\documentclass[usenatbib]{mn2e}
\usepackage{bm,graphicx,graphics}
\begin{document}

\title{Aberration by gravitational lenses in motion}

\author[S. Frittelli]
       {Simonetta Frittelli\\
    Department of Physics, Duquesne University,
        Pittsburgh, PA 15282}
\maketitle

\date{\today}

\begin{abstract}

It is known that a fully relativistic integration of the null geodesics of a
weak perturbation of flat spacetime leads to a correction of order $v/c$ to the
bending angle and time delay due to a gravitational lens in slow motion with
small acceleration.  The existence of the $v/c$ correction was verified by the
VLBI experiment of the bending of light by Jupiter on September 8, 2002. Here
the $v/c$ correction is interpreted by means of standard aberration of light in
an optically active medium with an effective index of refraction induced by the
gravitational field of a lens in motion.

\end{abstract}



\section{Introduction}

On September 8, 2002, the close alignment of Jupiter with quasar J0842+1835
provided the opportunity for an extraordinary measurement that consisted of the
delay between the observations of the quasar as recorded by two radio antennae.
The observation was conducted by the National Radio Astronomical Observatory
(USA) and the Max Planck Institute for Radio Astronomy. The time delay refers
to the difference in arrival time of two lightrays leaving the quasar
simultaneously at a single emission event. The observation ~\citep{fomalont}
verified the existence of a correction of order $v/c$ to the time delay of a
gravitational lens in motion: the fact that the time delay of light signals by
a point lens in motion carries a factor of $(1+v/c)$ with respect to the
Shapiro time delay --which, in turn, represents the additional time imposed on
a light signal due to the slow down by the gravitational field of a point mass
at rest. Here $v$ is the component of the velocity of the deflector along the
line of sight. 

The $v/c$ correction of first order to the bending angle was predicted (to the
best of my knowledge) by \cite{pyne1}, and subsequently confirmed by
\cite{kopeikin99}, \cite{fermatkling} and \cite{mnras}. The recent
observational verification strongly suggests that such a correction needs to be
incorporated into the fitting of gravitational lensing events. The size of the
correction is extremely small and may be negligible at cosmological scales.
Nevertheless, the effect may not only be important but even controllable in
microlensing events such as nearby stars with high proper motions acting as
lenses for distant stars in the Milky Way or in the Magellanic Clouds
\citep{mnras,paczynski96}.  

Here it is shown that the $v/c$ effect admits a natural interpretation in the
standard framework of gravitational lensing. In this framework, the bending of
light is interpreted in terms of light propagating within a medium with an
effective index of refraction, rather than in terms of a metric
field~\citep{pettersbook,EFS}.  Formally, the bending is analogous to that
induced by a piece of glass with a non-uniform index of refraction. The value
of this interpretation resides in its potential to integrate the $v/c$ effect
with other standard practices in gravitational lensing.

\section{Aberration by gravitational lenses} \label{sec:2}

It is well known~\citep{bergmann} that the apparent angular position of distant
stars wobbles as the Earth turns in its orbit around the Sun. The wobble is the
aberration caused by the observation of a lightray in two different frames: the
Sun's rest frame, and the inertial frame instantaneously attached to the Earth
at the moment of observation. In the Sun's rest frame at the observation event, the direction of a
lightray and the velocity of Earth define a plane where coordinates $(x,y)$ can
be defined so that the incoming lightray makes an angle $\theta$ with the
negative $y-$axis, and so that the Earth's 3-velocity has components
$\vec{v}=(v^x,v^y,0)$. In the Earth's frame the coordinates are $(x',y',z',ct')$
with
\begin{eqnarray}
\vec{r}' &=& \vec{r} -\vec{v}t +
\vec{v}(\gamma-1)\left(\frac{\vec{v}\cdot\vec{r}}{v^2} -t\right)\\
t'       &=& \gamma\left(t- \frac{\vec{v}\cdot\vec{r}}{c^2}\right)
\end{eqnarray}

\noindent where $\gamma=(1-(v/c)^2)^{-1/2}$. With the convention that $ds^2 =
\eta_{ab}dx^adx^b= -d(ct)^2+dx^2+dy^2+dz^2$, the Lorentz transformation applied
to the null vector $k^a$ in the Sun's rest frame given by
\[
k^a = (k^x,k^y,0,1)
\]

\noindent where $(k^x)^2 + (k^y)^2 =1$ yields the null vector $k'^a =
k^b\partial_bx'^a$ as seen in the Earth's frame of reference, the spatial part
of which being 
\[ \vec{k}'= \vec{k} -\frac{\vec{v}}{c} +(\gamma-1)\frac{\vec{v}}{c}
	\left(\frac{c\vec{k}\cdot \vec{v}}{v^2}-1\right).
\]

\noindent The observed angle with the negative $y-$axis on Earth is
$\tan\theta' \equiv -k'^x/k'^y$. We are only interested in the leading
correction of order $v/c$, so we may neglect the term proportional to
$(\gamma-1)$ to get
\begin{eqnarray}
\frac{k'^x}{k'^y} &=& \frac{k^x-v^x/c}{k^y-v^y/c} +O((v/c)^2)\\
		  &=& \left(1+\frac{v^y}{ck^y}\right)\frac{k^x}{k^y}
			- \frac{v^x}{ck^y} +O((v/c)^2).
\end{eqnarray}

\noindent By definition we have $k^x = \sin\theta$ and $k^y = -\cos\theta$
so this relationship reads
\[
\tan\theta' = \left(1-\frac{v^y}{c\cos\theta}\right)\tan\theta
			- \frac{v^x}{c\cos\theta},
\]

\noindent that is: the standard aberration formula for angles $\theta$ of any
size between $0$ and $\pi/2$, correct up to second order in $v/c$~\citep{bergmann}. For small
angles $\theta$ this simplifies to 

\begin{equation}\label{aberration}
\theta' = \left(1-\frac{v^y}{c}\right)\theta
			- \frac{v^x}{c},
\end{equation}

\noindent which shows that the two components of the Earth's velocity have
completely different aberration effects. The ``longitudinal'' component, $v^y$,
which at this level of accuracy is indistinguishable from the component of the
velocity along the line of sight, affects the apparent angle as a prefactor,
whereas the ``transverse'' component $v^x$ simply adds a fixed shift to all
small angles. 

Next let's consider gravitational lensing in its simplest form, that is: as the
propagation of a lightray within an optically active medium of index of
refraction $n$. The Newtonian potential $U=-GM/r$ of a point mass at rest
effectively slows down a lightray, inducing an effective index of refraction
$n=1-2U/c^2$ (by Eq.~(4.16) of \cite{EFS}).  By Fermat's principle, the path of the
lightray follows the direction of the unit vector $\hat{k}$ that minimizes the
travel time (see, for instance,~\cite{rossi}):
\begin{equation}
	\frac{d\hat{k}}{d\ell} 
      = \hat{k}\times \left[ \frac{\nabla n}{n}\times \hat{k}\right],
\end{equation}

\noindent where $d\ell$ is the Euclidean element of length along the
path.  Since $\hat{k}\times( \nabla n\times \hat{k}) = \nabla n -
(\hat{k}\cdot\nabla n )\hat{k}$, we have, equivalently,
\begin{equation}
	\frac{d\hat{k}}{d\ell} 
      = \frac{\nabla_\perp n}{n},
\end{equation}
			    
\noindent where the symbol $\perp$ indicates the local plane perpendicular
to $\hat{k}$. From emission to reception, the lightray changes direction by
\[
\vec{\alpha} \equiv \hat{k}_{in} - \hat{k}_{out},
\]

\noindent where the sublabels indicate the directions in which the
lightray effectively enters  and leaves the gravitational field of the
point mass. Since the path of the lightray stays on a plane, we can
restrict attention to the magnitude of $\vec{\alpha}$ which thus represents the change
between the angles that the incoming and exiting lightrays make with the
negative $y-$axis:  
\[
\alpha = \theta_{in} - \theta_{out}.
\]

\noindent In the case of a point mass at rest, the bending angle is 
\begin{equation}
\alpha=\frac{4GM}{c^2 \xi}
\end{equation}

\noindent where $\xi$ is the distance of closest approach of the lightray to the
point mass. 
This would be the bending of light that a deflector would induce on a
lightray, in the deflector's rest frame, since, during the fast transit of a
lightray, the deflector is well approximated as an inertial frame.  But since
the deflector is in motion with velocity $\vec{v}_d$ relative to the observer, 
which is approximately constant for the duration of the transit of the lightray
in its vicinity,  then from the point of view of the observer both angles
$\theta_{in},\theta_{out}$ --as well as the entire path during the transit--
are aberrated according to (\ref{aberration}) where $\vec{v} = -\vec{v}_d$. In
the rest frame of the observer we have
\begin{eqnarray}\label{finalangle}
\alpha' &\equiv& \theta'_{in} - \theta'_{out}\nonumber\\
	&=& \left( 1 +\frac{v^y_d}{c}\right)(\theta_{in} -\theta_{out})\nonumber\\
        &=& \left( 1 +\frac{v^y_d}{c}\right) \alpha .
\end{eqnarray}

\noindent Therefore the bending angle by a deflector moving with speed $v$
along the line of sight carries a prefactor of $(1+v/c)$ with respect to the
bending angle by the same deflector at rest.  The component of the velocity of
the deflector transverse to the line of sight does not affect the bending angle
(it affects both the incoming and exiting directions by the same shift, hence
not their difference). 

Now let's see what this implies for the time delay by a deflector in motion
with respect to the same deflector at rest.  By definition we have
\[
\vec{\alpha}= -\int \frac{d\hat{k}}{d\ell} d\ell
\]

\noindent Since $n$ differs from 1 only in terms that are of the order of
the Newtonian potential (the mass), then 
\[ \frac{d\hat{k}}{d\ell} = \nabla_\perp (n-1) + O(U^2),
\]

\noindent and 
\begin{equation}
\int \frac{d\hat{k}}{d\ell} d\ell= \nabla_\perp \int (n-1) d\ell
			    = \nabla_\perp (c\Delta t),
\end{equation}

\noindent where $\Delta t$ is the Shapiro time delay by the lens at rest:
\begin{equation}
\Delta t = -\frac{4GM}{c^3}\ln\xi.
\end{equation} 

\noindent So the bending angle and the gravitational time delay are related by a
gradient, that is, the time delay acts as a potential for the deflection:
\begin{equation}
	\vec{\alpha} = - \nabla_\perp (c\Delta t).
\end{equation}

The same is true if the lens is moving so far as the motion of the lens
is encoded in the corresponding index of refraction $n'$ resulting in a
corresponding
gravitational time delay $\Delta t'$, that is:
\begin{equation}
	\vec{\alpha}' = - \nabla_\perp (c\Delta t')
\end{equation}

\noindent It follows that 
\begin{equation}\label{finaldelay}
	\Delta t ' = \left( 1 +\frac{v^y}{c}\right) \Delta t,
\end{equation}

\noindent up to an additive constant. In this picture of
gravitational lensing, the time delay by a deflector in motion with speed $v$
along the line of sight carries a prefactor of $(1+v/c)$ with respect to the
standard Shapiro time delay, essentially due to the constancy of the speed of
light. 

One may question whether general relativity was used at all in this simplified
scheme. The answer is yes, in two ways. First, the factor of $2$ in the
effective index of  refraction $n=1-2U/c^2$ is a direct effect of the notion
that light travels along null geodesics of the metric $ds^2=-(1+2U/c^2)c^2dt^2
+(1-2U/c^2) d\vec{x}\cdot d\vec{x}$. In Newtonian fashion, on the contrary,
conservation of energy (per unit mass) for an orbit with instantaneous speed
$u$ that tends asymptotically to speed $c$ leads to $c^2/2 = u^2/2 + U$, which
yields an effective index of refraction $c/u\equiv n_{Newt} = 1-U/c^2 +O(U^2)$.
This difference in index of refraction leads to a difference of a factor of 2
in the bending angle, and, as is well known, constitutes the main and only
difference between Newtonian and relativistic bending of light in the weak
field regime of a point mass at rest, being the object of the classical test of
1919. This is discussed, for instance, in Section 1.1.1 of \cite{EFS}.

But the second way in which general relativity is used in the present
derivation of the $v/c$ effect is in the fact that the gravitational physics
involved is Lorentz invariant. The physics resides in the index of refraction
$n$.  It is known that under a Lorentz transformation, the index of refraction
of an optical medium transforms as $n^{-1}=n_s^{-1}-(v/c)(1-n_s^{-2})$ where
$n_s$ is the index of refraction of the same medium at rest, as demonstrated by
Fizeau in the classic experiment of light traversing a moving fluid in a pipe
\citep{bergmann}. In gravitational terms, where $n_s=1-2U/c^2$, this translates
directly to $n=1-2U/c^2-4(v/c)U/c^2$. This index of refraction is entirely
consistent with the metric $ds^2=-(1+2U/c^2)c^2dt^2 -8 (\vec{V}/c^2)\cdot
d\vec{x} \,cdt +(1-2U/c^2) d\vec{x}\cdot d\vec{x}$ with $\vec{V} \equiv U
\vec{v}/c$ \citep{EFS}, which is related by a Lorentz transformation to the
metric $ds_s^2=-(1+2U/c^2)c^2dt^2 +(1-2U/c^2) d\vec{x}\cdot d\vec{x}$ of a body
at rest.  Because the linearized Einstein equations are wave equations, which
are Lorentz invariant, \textit{both\/} metrics are solutions (for exactly the
same reason that the electric field of a uniformly moving charge coincides with
the Lorentz-transformed field of a charge at rest). The verification of the
$v/c$ effect is thus a direct check of the Lorentz invariance of the linearized
Einstein equations, in a way that the Eddington solar eclipse expedition was
not. The Jupiter experiment is, thus, nothing short of remarkable. 

The significance of the Jupiter experiment has been overshadowed by a
controversy as to whether or not the experiment directly measured the ``speed
of gravity'' (hence beating the gravitational wave detectors in the race for
the first direct detection of gravitational waves).  Because the experiment
actually found the field equations of general relativity to be Lorentz
invariant in the weak limit, there is hardly any question that the experiment
actually measured the speed of the characteristics of the field equations and
found it consistent with the prediction: the speed of light. What this means is
that, since the characteristics are lightlike,  when gravitational waves occur,
they must travel at the speed of light. This does provide a foundation to
interpret $c$ in the prefactor of $(1+v/c)$ in the time delay as the ``speed of
gravity''. But this does not mean that gravitational waves emitted by Jupiter
affected the measurement. The effect of gravitational waves in the experiment
is practically negligible, as thoroughly demonstrated by \cite{will03}. The
experiment did not directly measure the gravitational waves emitted by Jupiter.
This is the foundation of Will's technical claim that $c$ in the prefactor of
$(1+v/c)$ is not the ``speed of gravity'' \citep{will03}. It would seem as if
semantics had been playing a part in fueling up the controversy. It is to be
hoped that the present interpretation of the effect by simple aberration will
help settle the significance of the Jupiter experiment and provide a strong
case for corrections to microlensing.

\section*{acknowledgments} 

I am indebted to Patrick Riggs for calling my attention to the Jupiter
experiment. This work was supported by the NSF
under grant No. PHY-0070624 to Duquesne University.


\end{document}